\documentclass[fleqn,twoside]{article}
\usepackage{amsmath}
\usepackage{espcrc2}

\usepackage{graphicx}

\newcommand{\half}{{\frac{1}{2}}}
\newcommand{\mbf}[1]{\mathbf{#1}}

\title{Light-Front Holography and Gauge/Gravity Duality: The Light Meson and Baryon Spectra}

\author{ Guy F. de T\'eramond\address{Universidad de Costa Rica, San Jos\'e, Costa Rica}
and Stanley J. Brodsky\address{SLAC National Accelerator Laboratory, 
Stanford University, Stanford, CA 94309, USA} 
}
 
\begin{document}

\begin{abstract}
Starting from the bound state Hamiltonian equation of motion in QCD, we derive  relativistic light-front wave equations in terms of an invariant impact variable $\zeta$ which measures the separation of the quark and gluonic constituents within the hadron at equal light-front time. These equations of motion in physical space-time are  equivalent to the equations of motion which describe the propagation of spin-$J$ modes in anti--de Sitter (AdS) space. 
Its eigenvalues give the hadronic spectrum, and its eigenmodes represent the probability amplitudes of the hadronic constituents at a given scale. 
An effective classical gravity 
description in a positive-sign dilaton background
 $\exp(+ \kappa^2 z^2)$
 is given for the phenomenologically successful soft-wall model which naturally encodes the internal structure of hadrons and their orbital angular momentum. Applications to the light meson and baryon spectrum are presented. 

\vspace{0.7pc}

\end{abstract}

\maketitle

\section{Introduction}

We have recently shown  a remarkable
connection between the description of hadronic modes in AdS space and
the Hamiltonian formulation of QCD in physical space-time quantized
on the light-front (LF) at fixed light-front time $\tau = t + z/c$, the time marked by the
front of a light wave,~\cite{Dirac:1949cp} rather than at instant time $t$, the ordinary time.
In fact, we can use a first semiclassical approximation  to transform the fixed 
LF time bound-state Hamiltonian equation  to  a corresponding wave equation in AdS space. 
To this end, 
we identify an invariant LF
coordinate $\zeta$ which allows the separation of the dynamics of quark and gluon binding from
the kinematics of constituent spin and internal orbital angular momentum.~\cite{deTeramond:2008ht}
The result is a single-variable LF relativistic
Schr\"odinger equation  which determines the spectrum
and LF wavefunctions (LFWFs) of hadrons for general spin and
orbital angular momentum. This LF wave equation serves as a semiclassical first 
approximation to QCD, and it is equivalent to the
equations of motion which describe the propagation of spin-$J$ modes
on  AdS. 
Remarkably, the AdS equations
yield the kinetic energy terms of  the partons inside a
hadron, whereas the interaction terms build confinement and correspond to
truncation of AdS space~\cite{deTeramond:2008ht} in an effective dual gravity  approximation.

In this note we briefly review  the LF quantization of QCD  and its
LF Fock representation. We derive a relativistic light-front Schr\"odinger equation from the Hamiltonian bound state equation in light-front QCD
following Ref. \cite{deTeramond:2008ht}.  The identification of orbital angular momentum of the constituents is a key element in our description of the internal structure of hadrons using holographic principles. In our approach  quark and gluon degrees of freedom are explicitly introduced in the gauge/gravity correspondence, in contrast with the usual
AdS/QCD framework~\cite{Erlich:2005qh,DaRold:2005zs} where axial and vector currents become the primary entities as in effective chiral theory.
We will also review some of the features of the ``hard-wall''~\cite{Polchinski:2001tt}  and ``soft-wall''~\cite{Karch:2006pv}   which provide an initial approximation to QCD.  In our approach
the holographic mapping is carried out in the  strongly coupled regime where QCD is almost conformal, 
corresponding to an infrared fixed-point.~\cite{Deur:2008rf}
Our analysis follows from recent developments in light-front
QCD~\cite{Brodsky:2003px,deTeramond:2005su,Brodsky:2006uqa,Brodsky:2007hb,Brodsky:2008pf} which have been inspired by the
AdS/CFT correspondence.~\cite{Maldacena:1997re}

\section{Light-Front Quantization of QCD}

One can  express the hadron four-momentum  generator $P =  (P^+, P^-,
\mbf{P}_{\!\perp})$, $P^\pm = P^0 \pm P^3$, in terms of the
dynamical fields, the Dirac field $\psi_+$, 
where
$\psi_\pm = \Lambda_\pm
\psi$, $\Lambda_\pm = \gamma^0 \gamma^\pm$, and the transverse field
$\mbf{A}_\perp$ in the $A^+ = 0$ gauge~\cite{Brodsky:1997de}
 \begin{eqnarray} \nonumber
P^-  \!\!&=&\!  \half  \! \int \! dx^- d^2 \mbf{x}_\perp  \, \bar \psi_+  \gamma^+
\frac{m^2 \!+ \left( i \mbf{\nabla}_{\! \perp} \right)^2}{ i \partial^+} \psi_+  \cr
& & ~~~~~~~~~~~~~~~~~~~~~~~~~~ + {\rm (interactions)}, \\
\nonumber
P^+ \!\!&=&\!   \int \! dx^- d^2 \mbf{x}_\perp \,
 \bar \psi_+ \gamma^+   i \partial^+ \psi_+, \\ \label{eq:P}
\mbf{P}_{\! \perp}  \!\!&=&\!  \half \int \! dx^- d^2 \mbf{x}_\perp \,
\bar \psi_+ \gamma^+   i \mbf{\nabla}_{\! \perp} \psi_+,
\end{eqnarray}
where the integrals are over the initial surface $x^+ = 0$, $x^\pm = x^0 \pm x^3$.
The operator $P^-$ generates LF time translations
$\left[\psi_+(x), P^-\right] = i \frac{\partial}{\partial x^+} \psi_+(x)$,
and the generators $P^+$ and $\mbf{P}_\perp$ are kinematical.
For simplicity we have omitted from (\ref{eq:P})
the contribution from the gluon field $\mbf{A}_\perp$.

The Dirac field operator is expanded as
\begin{multline} \label{eq:psiop}
\psi_+(x^- \!,\mbf{x}_\perp)_\alpha = \sum_\lambda \int_{q^+ > 0} \frac{d q^+}{\sqrt{ 2
 q^+}}
\frac{d^2 \mbf{q}_\perp}{ (2 \pi)^3} \\ \times
\left[b_\lambda (q)
u_\alpha(q,\lambda) e^{-i q \cdot x}  + 
 d_\lambda (q)^\dagger
v_\alpha(q,\lambda) e^{i q \cdot x}\right],
\end{multline}
with $u$ and $v$ LF spinors. Similar expansion follows for $\mbf{A}_\perp$.
Using  LF
commutation relations 
$\left\{b(q), b^\dagger(q')\right\} = (2 \pi)^3 \,\delta (q^+ \! - {q'}^+)
\delta^{(2)} \! \left(\mbf{q}_\perp\! - \mbf{q}'_\perp\right)$,
one has
\begin{multline} \label{eq:Pm}
P^- \! =  \sum_\lambda \! \int \!  \frac{dq^+ d^2 \mbf{q}_\perp}{(2 \pi)^3 }   \,
\frac{m^2 \! + \mbf{q}_\perp^2}{q^+}  \,
 b_\lambda^\dagger(q) b_\lambda(q) \\ + { \rm (interactions)}.
\end{multline}
 The LF time evolution operator
$P^-$ is conveniently written as a term which represents the sum of the kinetic energy of all the partons plus a sum of all the interaction terms.

It is convenient to define a LF Lorentz invariant Hamiltonian
$H_{LF}= P_\mu P^\mu = P^-P^+  \! - \mbf{P}^2_\perp$ with eigenstates
$\vert \psi_H(P^+, \mbf{P}_\perp, S_z )\rangle$
and eigenmass  $\mathcal{M}_H^2$, the mass spectrum of the color-singlet states
of QCD~\cite{Brodsky:1997de}
\begin{equation} \label{eq:HLF}
H_{LF} \vert \psi_H\rangle = {\cal M}^2_H \vert \psi_H \rangle.
\end{equation}
A state $\vert \psi_H \rangle$ is an expansion 
in multi-particle Fock states
$\vert n \rangle $ of the free LF Hamiltonian:
~$\vert \psi_H \rangle = \sum_n \psi_{n/H} \vert n \rangle$, where
a one parton state is $\vert q \rangle = \sqrt{2 q^+} \,b^\dagger(q) \vert 0 \rangle$.
The Fock components $\psi_{n/H}(x_i, {\mathbf{k}_{\perp i}}, \lambda_i^z)$
are independent of  $P^+$ and $\mbf{P}_{\! \perp}$
and depend only on relative partonic coordinates:
the momentum fraction
 $x_i = k^+_i/P^+$, the transverse momentum  ${\mathbf{k}_{\perp i}}$ and spin
 component $\lambda_i^z$. Momentum conservation requires
 $\sum_{i=1}^n x_i = 1$ and
 $\sum_{i=1}^n \mathbf{k}_{\perp i}=0$.
The LFWFs $\psi_{n/H}$ provide a
{\it frame-independent } representation of a hadron which relates its quark
and gluon degrees of freedom to their asymptotic hadronic state.

\section{Light-Front Holography}

We can compute $\mathcal{M}^2$ from the hadronic matrix element
\begin{equation}
\langle \psi_H(P') \vert H_{LF}\vert\psi_H(P) \rangle  = 
\mathcal{M}_H^2  \langle \psi_H(P' ) \vert\psi_H(P) \rangle,
\end{equation}
expanding the initial and final hadronic states in terms of its Fock components. The computation is  simplified in the 
frame $P = \big(P^+, M^2/P^+, \vec{0}_\perp \big)$ where $H_{LF} =  P^+ P^-$.
We find
 \begin{multline} \label{eq:M}
 \mathcal{M}_H^2  =  \sum_n  \! \int \! \big[d x_i\big]  \! \left[d^2 \mbf{k}_{\perp i}\right]   
 \sum_q \Big(\frac{ \mbf{k}_{\perp q}^2 + m_q^2}{x_q} \Big)  \\ \times
 \left\vert \psi_{n/H} (x_i, \mbf{k}_{\perp i}) \right \vert^2  + {\rm (interactions)} ,
 \end{multline}
plus similar terms for antiquarks and gluons ($m_g = 0)$. The integrals in (\ref{eq:M}) are over
the internal coordinates of the $n$ constituents for each Fock state 
\begin{equation}
\int \big[d x_i\big] \equiv
\prod_{i=1}^n \int dx_i \,\delta \Bigl(1 - \sum_{j=1}^n x_j\Bigr) ,
\end{equation}
\vspace{-10pt}
\begin{equation}
\int \left[d^2 \mbf{k}_{\perp i}\right] \equiv \prod_{i=1}^n \int
\frac{d^2 \mbf{k}_{\perp i}}{2 (2\pi)^3} \, 16 \pi^3 \,
\delta^{(2)} \negthinspace\Bigl(\sum_{j=1}^n\mbf{k}_{\perp j}\Bigr),
\end{equation}
with phase space normalization
\begin{equation}
\sum_n  \int \big[d x_i\big] \left[d^2 \mbf{k}_{\perp i}\right]
\,\left\vert \psi_{n/H}(x_i, \mbf{k}_{\perp i}) \right\vert^2 = 1.
\end{equation}
The spin indices have been suppressed.
The LFWF $\psi_n(x_i, \mathbf{k}_{\perp i})$ can be expanded in terms of  $n-1$ independent
position coordinates $\mathbf{b}_{\perp j}$,  $j = 1,2,\dots,n-1$, 
conjugate to the relative coordinates $\mbf{k}_{\perp i}$
\begin{multline} \label{eq:LFWFb}
\psi_n(x_j, \mathbf{k}_{\perp j}) =  (4 \pi)^{(n-1)/2} 
\prod_{j=1}^{n-1}\int d^2 \mbf{b}_{\perp j}  \\
\exp{\Big(i \sum_{k=1}^{n-1} \mathbf{b}_{\perp k} \cdot \mbf{k}_{\perp k}\Big)} \,
{\psi}_n(x_j, \mathbf{b}_{\perp j}),
\end{multline}
where $\sum_{i = 1}^n \mbf{b}_{\perp i} = 0$.  We can also express (\ref{eq:M})
in terms of the internal coordinates $\mbf{b}_{\perp j}$ with the result
\begin{multline}   
 \mathcal{M}_H^2  =  \sum_n  \prod_{j=1}^{n-1} \int d x_j \, d^2 \mbf{b}_{\perp j} \,
\psi_n^*(x_j, \mbf{b}_{\perp j})  \\
 \sum_q   \left(\frac{ \mbf{- \nabla}_{ \mbf{b}_{\perp q}}^2  \! + m_q^2 }{x_q} \right) 
 \psi_n(x_j, \mbf{b}_{\perp j}) \\
  + {\rm (interactions)} . \label{eq:MKb}
 \end{multline}
The normalization is defined by
\vspace{-4pt}
\begin{equation}  \label{eq:Normb}
\sum_n  \prod_{j=1}^{n-1} \int d x_j d^2 \mathbf{b}_{\perp j}
\left \vert \psi_{n/H}(x_j, \mathbf{b}_{\perp j})\right\vert^2 = 1.
\end{equation}

To simplify the discussion we will consider a two-parton hadronic bound state.  In the limit
of zero quark mass
$m_q \to 0$
\begin{multline}  \label{eq:Mb}
\mathcal{M}^2  =  \int_0^1 \! \frac{d x}{x(1-x)} \int  \! d^2 \mbf{b}_\perp  \,
  \psi^*(x, \mbf{b}_\perp) \\ \times
  \left( - \mbf{\nabla}_{ {\mbf{b}}_{\perp}}^2\right)
  \psi(x, \mbf{b}_\perp) +   {\rm (interactions)}.
 \end{multline}

 The functional dependence  for a given Fock state is
given in terms of the invariant mass
\begin{equation}
 \mathcal{M}_n^2  = \Big( \sum_{a=1}^n k_a^\mu\Big)^2 = \sum_a \frac{\mbf{k}_{\perp a}^2 +  m_a^2}{x_a}
 \to \frac{\mbf{k}_\perp^2}{x(1-x)} \,,
 \end{equation}
 the measure of  the off-energy shell of the bound state,
 $\mathcal{M}^2 \! - \! \mathcal{M}_n^2$.
 Similarly in impact space the relevant variable for a two-parton state is  $\zeta^2= x(1-x)\mbf{b}_\perp^2$.
Thus, to first approximation  LF dynamics  depend only on the boost invariant variable
$\mathcal{M}_n$ or $\zeta,$
and hadronic properties are encoded in the hadronic mode $\phi(\zeta)$ from the relation
\begin{equation} \label{eq:psiphi}
\psi(x,\zeta, \varphi) = e^{i M \varphi} X(x) \frac{\phi(\zeta)}{\sqrt{2 \pi \zeta}} ,
\end{equation}
thus factoring out the angular dependence $\varphi$ and the longitudinal, $X(x)$, and transverse mode $\phi(\zeta)$
with normalization $ \langle\phi\vert\phi\rangle = \int \! d \zeta \,
 \vert \langle \zeta \vert \phi\rangle\vert^2 = 1$.
 
We can write the Laplacian operator in (\ref{eq:Mb}) in circular cylindrical coordinates $(\zeta, \varphi)$
and factor out the angular dependence of the
modes in terms of the $SO(2)$ Casimir representation $L^2$ of orbital angular momentum in the
transverse plane. Using  (\ref{eq:psiphi}) we find~\cite{deTeramond:2008ht}
\begin{multline} \label{eq:KV}  
\mathcal{M}^2   =  \int \! d\zeta \, \phi^*(\zeta) \sqrt{\zeta}
\left( -\frac{d^2}{d\zeta^2} -\frac{1}{\zeta} \frac{d}{d\zeta}
+ \frac{L^2}{\zeta^2}\right)
\frac{\phi(\zeta)}{\sqrt{\zeta}}   \\
+ \int \! d\zeta \, \phi^*(\zeta) U(\zeta) \phi(\zeta) ,
\end{multline}
where all the complexity of the interaction terms in the QCD Lagrangian is summed up in the effective potential $U(\zeta)$.
The LF eigenvalue equation $H_{LF} \vert \phi \rangle  =  \mathcal{M}^2 \vert \phi \rangle$
is thus a light-front  wave equation for $\phi$
\begin{equation} \label{eq:QCDLFWE}
\left(-\frac{d^2}{d\zeta^2}
- \frac{1 - 4L^2}{4\zeta^2} + U(\zeta) \right)
\phi(\zeta) = \mathcal{M}^2 \phi(\zeta),
\end{equation}
an effective single-variable light-front Schr\"odinger equation which is
relativistic, covariant and analytically tractable. 

It is important to notice that in the light-front the $SO(2)$ Casimir for orbital angular momentum $L^2$
is a kinematical quantity, in contrast with the usual $SO(3)$ Casimir $\ell(\ell+1)$ from non-relativistic physics which is
rotational, but not boost invariant. Using (\ref{eq:MKb}) one can readily
generalize the equations to allow for the kinetic energy of massive
quarks.~\cite{Brodsky:2008pg}  In this case, however,
the longitudinal mode $X(x)$ does not decouple from the effective LF bound-state equations.

As the simplest example, we consider a bag-like model
where  partons are free inside the hadron
and the interaction terms effectively build confinement. The effective potential is a hard wall:
$U(\zeta) = 0$ if  $\zeta \le 1/\Lambda_{\rm QCD}$ and
 $U(\zeta) = \infty$ if $\zeta > 1/\Lambda_{\rm QCD}$,
 where boundary conditions are imposed on the
 boost invariant variable $\zeta$ at fixed light-front time.   If $L^2 \ge 0$ the LF Hamiltonian is positive definite
 $\langle \phi \vert H_{LF} \vert \phi \rangle \ge 0$ and thus $\mathcal M^2 \ge 0$.
 If $L^2 < 0$ the bound state equation is unbounded from below and the particle
 ``falls towards the center''. The critical value corresponds to $L=0$.
  The mode spectrum  follows from the boundary conditions
 $\phi \! \left(\zeta = 1/\Lambda_{\rm QCD}\right) = 0$, and is given in
 terms of the roots of Bessel functions: $\mathcal{M}_{L,k} = \beta_{L, k} \Lambda_{\rm QCD}$.
 Upon the substitution $\Phi(\zeta) \sim \zeta^{3/2} \phi(\zeta)$, $\zeta \to z$
 we find
 \begin{equation} \label{eq:eomPhiJz}
\left[ z^2 \partial_z^2 - 3 z \, \partial_z + z^2 \mathcal{M}^2
\!  -  (\mu R)^2 \right] \!  \Phi_J  = 0,
\end{equation}
 the wave equation which describes the propagation of a scalar mode in a fixed AdS$_5$ background with AdS radius $R$.
 The five dimensional mass $\mu$ is related to the orbital angular momentum of the hadronic bound state by
 $(\mu R)^2 = - 4 + L^2$. The quantum mechanical stability $L^2 >0$ is thus equivalent to the
 Breitenlohner-Freedman stability bound in AdS.~\cite{Breitenlohner:1982jf}
The scaling dimensions are $\Delta = 2 + L$ independent of $J$ in agreement with the
twist scaling dimension of a two parton bound state in QCD. Higher spin-$J$ wave equations
are obtained by shifting dimensions: $\Phi_J(z) = (z/R)^{-J} \Phi(z)$.~\cite{deTeramond:2008ht}

The hard-wall LF model discussed here is 
equivalent to the hard wall model of
 Ref.~\cite{Polchinski:2001tt}.   The variable $\zeta$,
 $0 \leq \zeta \leq \Lambda_{\rm QCD}^{-1}$,  represents the 
 invariant separation between pointlike constituents and is also
 the holographic variable $z$ in AdS, thus we can identify $\zeta = z$.
 Likewise a two-dimensional  oscillator with
 effective potential  $ U(z) = \kappa^4 z^2 + 2 \kappa^2(L+S-1)$ is similar to the soft-wall model of
 Ref.~\cite{Karch:2006pv} which reproduce the usual linear Regge trajectories, where $L$ is the internal
 orbital angular momentum and $S$ is the internal spin.
 As we will show below, the soft-wall discussed here correspond to a positive sign
 dilaton, and higher-spin solutions follow from shifting dimensions: $\Phi_J(z) = (z/R)^{-J} \Phi(z)$.

 The resulting mass spectra  for mesons  at zero quark mass is
${\cal M}^2 = 4 \kappa^2 (n + L +S/2)$. Thus one can compute the hadron spectrum by simply adding  $4 \kappa^2$ for a unit change in the radial quantum number, $4 \kappa^2$ for a change in one unit in the orbital quantum number and $2 \kappa^2$ for a change of one unit of spin to the ground state value of $\mathcal{M}^2$. Remarkably, the same rule holds for baryons as shown below.

\begin{figure}[!]
\begin{center}
\includegraphics[angle=0,width=7.4cm]{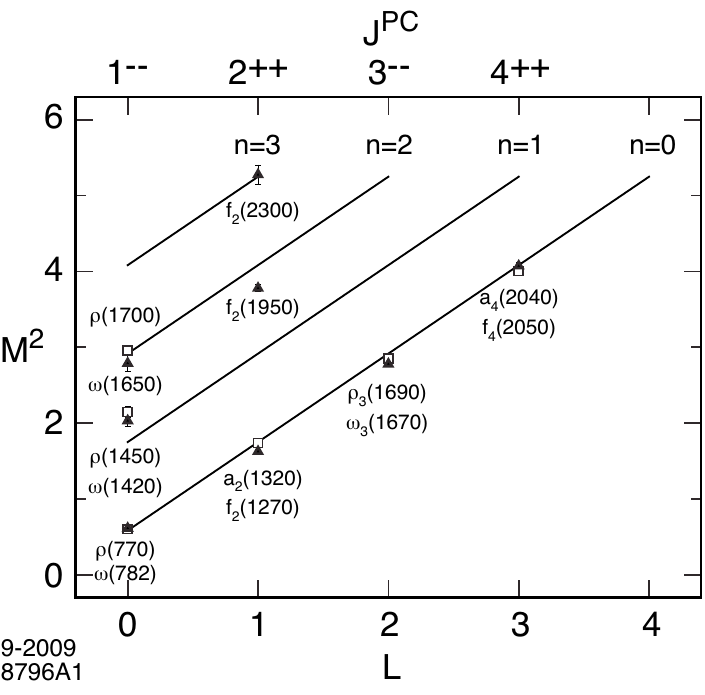}
\caption{Regge trajectories for the  $I\!=\!1$ $\rho$-meson
 and the $I\!=\!0$  $\omega$-meson families for $\kappa= 0.54$ GeV.}
\label{VM}
\end{center}
\end{figure}

Individual hadron states can be identified by their interpolating operator at $z\to 0.$  For example, the pseudoscalar meson interpolating operator
$\mathcal{O}_{2+L}= \bar q \gamma_5 D_{\{\ell_1} \cdots D_{\ell_m\}} q$, 
written in terms of the symmetrized product of covariant
derivatives $D$ with total internal  orbital
momentum $L = \sum_{i=1}^m \ell_i$, is a twist-two, dimension $3 + L$ operator
with scaling behavior determined by its twist-dimension $ 2 + L$. Likewise
the vector-meson operator
$\mathcal{O}_{2+L}^\mu = \bar q \gamma^\mu D_{\{\ell_1} \cdots D_{\ell_m\}} q$
has scaling dimension $\Delta=2 + L$.  The scaling behavior of the scalar and vector AdS modes $\Phi(z) \sim z^\Delta$ at $z \to 0$  is precisely the scaling required to match the scaling dimension of the local pseudoscalar and vector-meson interpolating operators.   
The spectral predictions for  light vector meson  states are compared with experimental data in the
Chew-Frautschi plot in Fig. \ref{VM} for the soft-wall model discussed here.

The twist dimension of the interpolating operator ensures dimensional counting rules for form factors and other hard exclusive processes, consistent with conformal invariance at short distances as well as the scaling expected from 
supersymmetry,~\cite{Craig:2009rk} 
since the scalar field,  
the spinor field and the gluon field G all have twist one.

\begin{figure}[!]
\begin{center}
\includegraphics[angle=0,width=7.4cm]{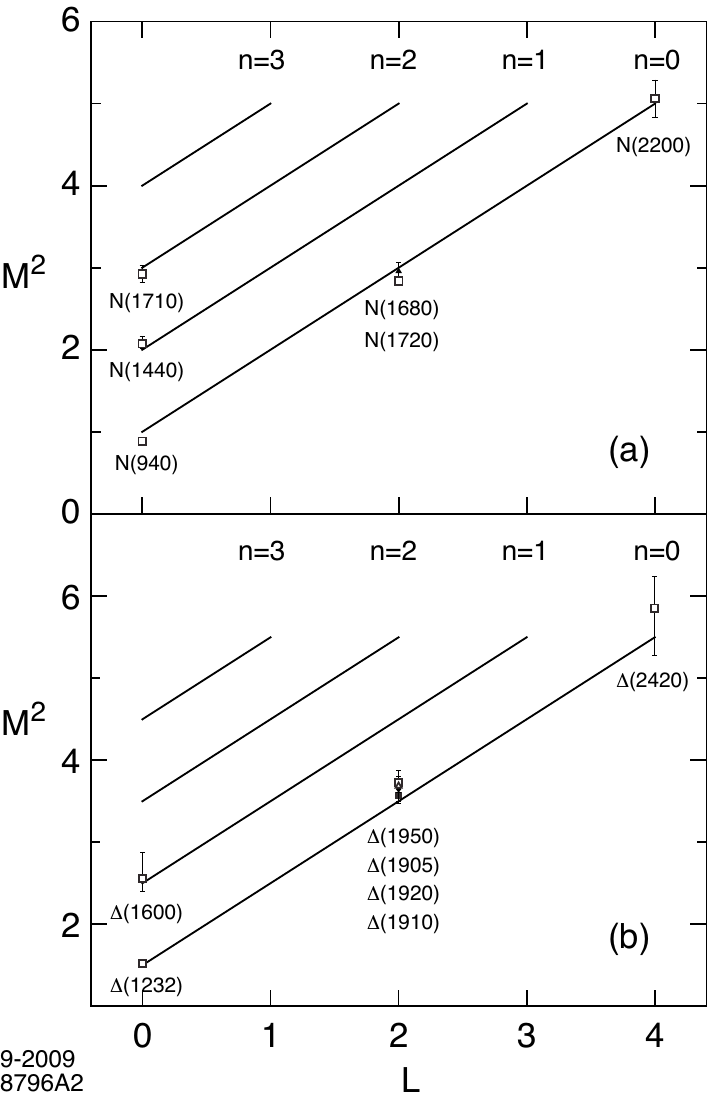}
\caption{{{\bf 56} Regge trajectories for  the  $N$ and $\Delta$ baryon families for $\kappa= 0.5$ GeV}
}
\label{Baryons}
\end{center}
\end{figure}

For baryons, the light-front wave equation is a linear equation 
determined by the LF transformation properties of spin 1/2 states. A linear confining potential 
$U(\zeta) \sim \kappa^2 \zeta$ in the LF Dirac
equation leads to linear Regge trajectories.~\cite{Brodsky:2008pg}   For fermionic modes the light-front matrix
Hamiltonian eigenvalue equation $D_{LF} \vert \psi \rangle = \mathcal{M} \vert \psi \rangle$, $H_{LF} = D_{LF}^2$, 
in a $2 \times 2$ spinor  component
representation is equivalent to the system of coupled linear equations
\begin{eqnarray} \label{eq:LFDirac} \nonumber
- \frac{d}{d\zeta} \psi_- -\frac{\nu+\half}{\zeta}\psi_- 
- \kappa^2 \zeta \psi_-&=&
\mathcal{M} \psi_+, \\ \label{eq:cD2k}
  \frac{d}{d\zeta} \psi_+ -\frac{\nu+\half}{\zeta}\psi_+ 
- \kappa^2 \zeta \psi_+ &=&
\mathcal{M} \psi_-. 
\end{eqnarray}
with eigenfunctions
\begin{eqnarray} \nonumber
\psi_+(\zeta) &\sim& z^{\frac{1}{2} + \nu} e^{-\kappa^2 \zeta^2/2}
  L_n^\nu(\kappa^2 \zeta^2) ,\\
\psi_-(\zeta) &\sim&  z^{\frac{3}{2} + \nu} e^{-\kappa^2 \zeta^2/2}
 L_n^{\nu+1}(\kappa^2 \zeta^2), 
\end{eqnarray}
and  eigenvalues
\begin{equation}
\mathcal{M}^2 = 4 \kappa^2 (n + \nu + 1) .
\end{equation}

The baryon interpolating operator
$ \mathcal{O}_{3 + L} =  \psi D_{\{\ell_1} \dots
 D_{\ell_q } \psi D_{\ell_{q+1}} \dots
 D_{\ell_m\}} \psi$,  $L = \sum_{i=1}^m \ell_i$, is a twist 3,  dimension $9/2 + L$ with scaling behavior given by its
 twist-dimension $3 + L$. We thus require $\nu = L+1$ to match the short distance scaling behavior. Higher spin fermionic modes are obtained by shifting dimensions for the fields as in the bosonic case. 
Thus, as in the meson sector,  the increase  in the 
mass squared for  higher baryonic states is
$\Delta n = 4 \kappa^2$, $\Delta L = 4 \kappa^2$ and $\Delta S = 2 \kappa^2$, 
relative to the lowest ground state,  the proton.

The predictions for the $\bf 56$-plet of light baryons under the $SU(6)$  flavor group are shown in Fig. \ref{Baryons}. As for the predictions for mesons in Fig. \ref{VM}, only confirmed PDG~\cite{Amsler:2008xx} states are shown. 
The Roper state $N(1440)$ and the $N(1710)$ are well accounted for in this model as the first  and second radial
states. Likewise the $\Delta(1660)$ corresponds to the first radial state of the $\Delta$ family. The model is  successful in explaining the important parity degeneracy observed in the light baryon spectrum, such as the $L\! =\!2$, $N(1680)\!-\!N(1720)$ pair and the $\Delta(1905), \Delta(1910), \Delta(1920), \Delta(1950)$ states which are degenerate 
within error bars. Parity degeneracy of baryons is also a property of the hard wall model, but radial states are not well described by this model.~\cite{deTeramond:2005su}
Recent work on the hadronic spectrum based on AdS/QCD models is given in~\cite{BoschiFilho:2005yh,Vega:2008af,dePaula:2008fp,Colangelo:2008us,Forkel:2008un,Ahn:2009px,dePaula:2009za,Sui:2009xe}.

\section{A Soft-Wall Model}

A nonconformal metric dual to a confining gauge theory is conveniently written as~\cite{Polchinski:2001tt}
\begin{equation}
ds^2 = \frac{R^2}{z^2} e^{2 A(z)} \left( \eta_{\mu \nu} dx^\mu dx^\nu - dz^2\right),
\end{equation}
where the warp factor $A(z) \to 0$ at small $z$ for geometries which are asymptotically AdS$_5$.
According to Sonnenscheim~\cite{Sonnenschein:2000qm} a background dual to a confining theory
should satisfy the conditions for the time-time metric component $g_{00}$
\begin{equation} \label{Scond}
\partial_z (g_{00}) \vert_{z=z_0} = 0 , ~~~~ g_{00} \vert_{z = z_0} \ne 0,
\end{equation}
to display Wilson loop area law for confinement of strings.

We consider a warp factor $2 A(z) = \pm \kappa^2 z^2$~\cite{Andreev:2006vy,Andreev:2006ct}
which, for our present discussion can be considered similar to the dilaton background introduced in Ref. \cite{Karch:2006pv}. 
The metric for the positive sign  warp factor $2 A(z) = + \kappa^2 z^2$ satisfy the conditions (\ref{Scond}) with $z_0 = 1/\sqrt{2} \kappa$. This type of solution was considered in~\cite{Andreev:2006ct} to derive a confining potential between heavy quarks.
The area law confining conditions (\ref{Scond}) are not obeyed for the negative sign warp factor, $A(z) = - \kappa^2 z^2$, the solution described in \cite{Karch:2006pv}.

We may also study the confinement properties of the warped metrics by following an object in AdS space as it falls to the infrared region by the effects of gravity. The gravitational potential energy for an object of mass $m$ in general relativity is
given in terms of  $g_{00}$
\begin{equation} \label{V}
V = mc^2 \sqrt{g_{00}} = mc^2 R \, \frac{e^A(z)}{z}.
\end{equation}
For the negative solution the potential decreases monotonically, and thus an object in AdS will fall to infinitely large 
values of $z$. For the positive solution, the potential is nonmonotonic and has an absolute minimum at $\bar z = 1/\kappa$.  
For large values
of $z$ the gravitational potential increases exponentially, thus confining any object in the modified AdS metrics to distances $\langle z \rangle \sim 1/\kappa$.  Indeed the potential (\ref{V}) for a positive sign factor $2 A(z) = + \kappa^2 z^2$ is similar to the confining potential depicted in Ref. \cite{Klebanov:2009zz}.
We will study here the positive sign solution, which
as discussed above, has an interesting physical motivation.~\cite{note1}

For practical reasons we introduce a dilaton background $\varphi(z)$ as in Ref.   \cite{Karch:2006pv}, instead of deforming the AdS$_5$ metrics, and write the action
\begin{equation}
S = \int \! d^d x \, dz  \,\sqrt{g} \,e^{\varphi(z) } \mathcal{L}, 
\end{equation}
in AdS$_{d+1}$ for the dilaton field $\varphi(z) = \pm \kappa^2 z^2$. For a scalar field the Lagrangian $\mathcal{L} = \half  \big( g^{\ell m} \partial_\ell \Phi \partial_m \Phi 
-  \mu^2  \Phi^2 \big)$
leads to the wave equations
 \begin{multline}  \label{WES}
 \big[z^2 \partial_z^2 - \left(d-1 \, \mp \, 2 \kappa^2 z^2 \right) z\,\partial_z + z^2 \mathcal{M}^2  \\
 - (\mu R)^2 \big] \Phi(z) = 0
 \end{multline} 
 with $(\mu R)^2 \ge -4$.  For $\varphi(z) = - \kappa^2 z^2$ there is no stable solution for the lowest
 value  $(\mu R)^2 = -4$. For  $\varphi(z) = \kappa^2 z^2$ the lowest possible solution corresponding to
 $(\mu R)^2 = -4$ is $\Phi(z) \sim z^2  e^{-\kappa^2 z^2} \!,$ with eigenvalue $\mathcal{M}^2 = 0$. This is
 a chiral symmetric bound state of two massless quarks with scaling dimension 2 and size 
 $\langle z^2 \rangle \sim 1/\kappa^2$, which we identify with the lowest state, the pion.

We define a spin-$J$ mode $\Phi_{\mu_1 \cdots \mu_J}$  with all  indices 
along 3+1 with shifted dimensions
$\Phi_J(z) = ( z/R)^{-J}  \Phi(z)$ and normalization
\begin{equation} 
R^{d-2J-1} \int_0^{\infty} \! \frac{dz}{z^{d-2J-1}} \, e^{\kappa^2 z^2} \Phi_J^2 (z) = 1.
\end{equation}

The shifted field $\Phi_J$ obeys the 
wave equation
\begin{multline} \label{WESJ}
\big[ z^2 \partial_z^2 - \left(d\! -\! 1 \!- \!2 J - 2 \kappa^2 z^2 \right) z \, \partial_z + z^2 \mathcal{M}^2
\\ -  (\mu R)^2 \big] \Phi_J  = 0
\end{multline}
which follows from (\ref{WES})
upon mass rescaling ~~$(\mu R)^2 \to (\mu R)^2 - J(d-J)$  and $\mathcal{M}^2 \to \mathcal{M}^2 - 2 J \kappa^2$.

Upon the substitution ~$z \! \to\! \zeta$  and  $\phi_J(\zeta)   \!  \sim \! \zeta^{-3/2 + J} e^{\kappa^2 \zeta^2 /2} \, \Phi_J(\zeta)$,
 we find for $d=4$ the QCD light-front wave equation
\begin{multline}  \label{LFWEJ}
\Big(-\frac{d^2}{d \zeta^2} - \frac{1-4 L^2}{4\zeta^2} 
+ \kappa^4 \zeta^2 + 2 \kappa^2(L + S - 1)  \Big) \phi_{\mu_1 \cdots \mu_J} \\
= \mathcal{M}^2 \phi_{\mu_1 \cdots \mu_J},
\end{multline}
where $J_z = L_z + S_z$ and $(\mu R)^2 = - (2-J)^2 + L^2$. 

Equation  (\ref{LFWEJ}) has eigenfunctions
\begin{equation}
\phi_n^L(\zeta) = \kappa^{1+L} \sqrt{\frac{2 n!}{(n\!+\!L\!)!}} \, \zeta^{1/2+L}
e^{- \kappa^2 \zeta^2/2} L^L_n(\kappa^2 \zeta^2) ,
\end{equation}
and eigenvalues 
\begin{equation}
\mathcal{M}_{n, L, S}^2 = 4 \kappa^2 \left(n + L + \frac{S}{2} \right). 
\end{equation}
We thus recover the potential  $U(z) = \kappa^4 \zeta^2 + 2 \kappa^2(L + S - 1)$ used above for the 
computation of the meson spectrum.

\section{Conclusions}

We have derived a connection between a semiclassical first approximation to QCD quantized on the light-front
and hadronic modes propagating on a fixed AdS background. This
duality leads to a remarkable Schr\"odinger-like equation in the AdS 
fifth dimension coordinate $z$ (\ref{eq:QCDLFWE}).
We have shown how the identical AdS wave equation can be derived in physical space time as an effective equation for valence quarks in LF quantized theory, where one identifies the AdS fifth dimension coordinate $z$ with the LF coordinate $\zeta$.  
We originally derived this correspondence using the identity between electromagnetic and gravitational form factors computed in AdS and LF theory~\cite{Brodsky:2006uqa,Brodsky:2007hb,Brodsky:2008pf}. 
Our derivation 
also shows that the mass $\mu$ is directly related to orbital angular momentum $L$ in physical space-time. The result is physically compelling and phenomenologically successful.

 We have shown how the soft-wall AdS/CFT model with a dilaton-modified AdS space leads to the
 potential $U(z) = \kappa^4 z^2 + 2 \kappa^2(L+S-1)$. This potential can be derived directly from the action in AdS space and corresponds to a dilaton  profile 
 $\exp( +\kappa^2 z^2)$,  with  opposite 
sign to that of Ref.  \cite{Karch:2006pv}.  Hadrons are identified by matching the power behavior of the hadronic amplitude at the AdS boundary at small $z$ to the twist of its interpolating operator at short distances $x^2 \to 0$, as required by the AdS/CFT dictionary. The twist corresponds to the dimension of fields appearing in chiral super-multiplets. The twist of a hadron equals the number of constituents.

The light-front  AdS/QCD equation provides 
remarkably
successful predictions for the light-quark meson and baryon spectra, as function of hadron spin, quark angular momentum, and radial quantum number.~\cite{note2} 
The pion is massless, corresponding to zero mass quarks, in agreement with chiral invariance arguments.
The predictions for form factors are remarkable successful. The predicted power law fall-off agrees with dimensional counting rules as required by conformal invariance at small $z$.~\cite{Brodsky:2007hb,Brodsky:2008pg}

Higher spin modes follow from shifting dimensions in the AdS wave equations.
In the hard-wall model the dependence is linear:  $\mathcal{M} \sim 2n + L$. 
However, in the soft-wall
model the standard Regge behavior is found $\mathcal{M}^2 \sim n +
L$.   Both models predict the same multiplicity of states for mesons
and baryons, which is observed experimentally.~\cite{Klempt:2007cp}
As in the Schr\"odinger equation, the semiclassical approximation to light-front QCD 
described in this paper does not account for particle
creation and absorption; it is thus expected to break down at short distances
where hard gluon exchange and quantum corrections become important. 
However, one can systematically
improve the semiclassical approximation, for example by introducing nonzero quark masses and short-range Coulomb
corrections to describe the dynamics of heavy and heavy-light quark systems.

\section*{Acknowledgments}
 Presented by GdT at Light Cone 2009: Relativistic Hadronic and Particle Physics, 8-13 July 2009, S\~ao Jos\'e dos Campos, Brazil.  We are grateful to Professor Tobias Frederico  and his colleagues at the Instituto Tecnol\'ogico de Aeron\'autica (ITA), 
 for their outstanding hospitality.
This research was supported by the Department
of Energy  contract DE--AC02--76SF00515.  SLAC-PUB-13779.

\section*{Added Note}
Following this article a preprint by Fen Zuo appeared (arXiv:0909.4240) where it is shown that the introduction of a positive dilaton profile is particularly suited for describing chiral symmetry breaking. In contrast with the original model~\cite{Karch:2006pv},  the expectation value of the scalar field associated with the quark mass and condensate does not blow up in the far infrared region of AdS.

\end{document}